\documentclass[%
 aps,
 pra,%
 amsmath,amssymb,
 groupedaddress,
 showpacs,
 showkeys,
preprint,
]{revtex4-1}

\usepackage{graphicx}
\usepackage{dcolumn}
\usepackage{bm}

\usepackage[version=3]{mhchem}

\usepackage{hyperref}

\usepackage{epstopdf} 

\begin{document}

\preprint{APS-KD2014}

\title[Model for the overall phase-space acceptance in a Zeeman decelerator]{Model for the overall phase-space acceptance in a Zeeman decelerator}

\author{Katrin Dulitz}
\affiliation{Department of Chemistry, University of Oxford, Chemistry Research Laboratory, 12 Mansfield Road, Oxford, OX1 3TA, United Kingdom}
\author{Nicolas Vanhaecke}
\affiliation{Fritz-Haber-Institut der Max-Planck-Gesellschaft, Faradayweg 4-6, 14195 Berlin, Germany}
\affiliation{Laboratoire Aim\'{e} Cotton, CNRS, Universit\'{e} Paris-Sud, ENS Cachan, 91405 Orsay, France}
\author{Timothy P. Softley}
\email{tim.softley@chem.ox.ac.uk.}
\affiliation{Department of Chemistry, University of Oxford, Chemistry Research Laboratory, 12 Mansfield Road, Oxford, OX1 3TA, United Kingdom}

\date{\today}

\begin{abstract}

We present a new formalism to calculate phase-space acceptance in a Zeeman decelerator. Using parameters closely mimicking previous Zeeman deceleration experiments, this approach reveals a hitherto unconsidered velocity dependence of the phase stability which we ascribe to the finite rise and fall times of the current pulses that generate the magnetic fields inside the deceleration coils. 
It is shown that changing the current switch-off times (characterized by the reduced position of the synchronous particle $\kappa_0$) as the sequence progresses, so as to maintain a constant mean acceleration per pulse, can lead to a constant phase stability and hence a beam with well-defined characteristics.  We also find that the time overlap between fields of adjacent coils has an influence on the phase-space acceptance.
Previous theoretical and experimental results \cite{Wiederkehr2010, Wiederkehr2011} suggested unfilled regions in phase space that influence particle transmission through the decelerator. Our model provides, for the first time, a means to directly identify the origin of these effects due to coupling between longitudinal and transverse dynamics. Since optimum phase stability is restricted to a rather small parameter range in terms of the reduced position of the synchronous particle, $\kappa_0$, only a limited range of final velocities can be attained using a given number of coils. We evaluate phase stability for different Zeeman deceleration sequences, and, by comparison with numerical three-dimensional particle trajectory simulations, we demonstrate that our model provides a valuable tool to find optimum parameter sets for improved Zeeman deceleration schemes. An acceleration-deceleration scheme is shown to be a useful approach to generating beams with well-defined properties for variable-energy collision experiments. More generally, the model provides significant physical insights applicable to other types of particle decelerators with finite rise and fall time fields.  

\end{abstract}

\pacs{Valid PACS appear here}

\maketitle

\section{\label{sec:introduction} Introduction}

The past decade has seen numerous advances in the field of cold molecules, both in the development of experimental techniques for the production of cold and ultracold molecular samples, and in their application to the chemical and physical sciences \cite{Doyle2004, Schnell2009, Bell2009, Carr2009, Stuhl2014}. The manipulation of molecular beams using inhomogeneous, time-varying electromagnetic fields in multistage Stark and Zeeman decelerators in particular \cite{Hogan2011, van2012, Narevicius2012} has found applications in spectroscopy \cite{Veldhoven2004, Hudson2006, Motsch2014} and the study of state-selected collision processes \cite{Gilijamse2006a, Sawyer2008a, Kirste2012, vonZastrow2014}.

Zeeman deceleration is a beam deceleration technique, in which the velocity of a supersonic beam is reduced by successively switching strong magnetic fields inside an array of solenoid coils \cite{Vanhaecke2007a, Narevicius2008} such that particles in low-field-seeking quantum states, whose Zeeman energy increases with the magnetic field, always experience a positive magnetic field gradient. Upon application of a deceleration pulse sequence, their kinetic energy is thus gradually converted into Zeeman energy. The operation principles of a Zeeman decelerator are very similar to those used for Stark deceleration, in which high electric fields are rapidly switched inside an electrode array \cite{Bethlem1999, Bethlem2000a}. Zeeman and Stark deceleration are complementary techniques \cite{Hogan2011, van2012}, since Stark decelerators are used to manipulate the motion of particles with an electric dipole moment, while Zeeman decelerators make it possible to decelerate beams of paramagnetic species.

Both Stark and Zeeman deceleration are based on the concept of phase stability which was originally devised for charged-particle acceleration in synchrotrons \cite{McMillan1945, Veksler1945}. Phase stability explains why particles within a limited range of  positions and velocities are kept together throughout the deceleration process. The characteristics of a decelerated beam, e.g., particle transmission and velocity spread, are widely governed by phase-space acceptance, and a thorough understanding of this concept is needed for the design of improved deceleration sequences and experimental setups. phase-space acceptance in a Stark decelerator was first described by Bethlem et al. \cite{Bethlem2000, Bethlem2002}. The model was extended by van de Meerakker et al. to explain the influence of the transverse motion on the deceleration process \cite{van2006a}. The group also demonstrated that the periodicity of the electric fields leads to additional phase-stable regions inside a Stark decelerator \cite{van2005b}. Based on these ideas, they built a Stark decelerator with an improved overall performance if operated in a higher-order mode ($s$~=~3) \cite{Scharfenberg2009}. Particle motion in a multistage Zeeman decelerator and a moving-trap Zeeman decelerator has thus far been pictured through one-dimensional models, and via numerical three-dimensional particle-trajectory simulations \cite{Wiederkehr2010, LavertOfir2011a}.

Here, we present a model that allows for a more general understanding of the longitudinal and transverse acceptance as well as the overall six-dimensional phase-space acceptance in a Zeeman decelerator. It thus provides a useful means to find conditions for the phase-stable operation of a Zeeman decelerator without the need of having to run large sets of trajectory simulations in a multi-parameter space. The output of our model suggests that, for a given switch-off position inside a solenoid coil, $\kappa_0$ (defined below), the phase-space acceptance in a multistage Zeeman decelerator is dependent on the particle velocity, and on the time overlap of the current pulses between neighboring coils. In order to remain in the same phase-stable region throughout the deceleration process, we suggest the use of an adaptive $\kappa_0$ which follows the change in the mean longitudinal acceleration as the particle velocity is decreased.

Using our model, we are able to explain the origin of unfilled regions in phase space that have been predicted by theoretical and experimental studies \cite{Wiederkehr2010, Wiederkehr2011}, and that we also obtain using numerical three-dimensional particle trajectory simulations. To overcome this decrease in phase stability, we assess the efficiency of two alternative schemes for Zeeman deceleration that are similar to those used for the switching of electric fields in Stark decelerator experiments \cite{Bethlem1999, van2005b, Scharfenberg2009}. Furthermore, we outline a new mode of operation for a Zeeman decelerator that is based upon the alternation between Zeeman acceleration and deceleration. This scheme only relies on changes in the computed deceleration pulse sequence thus facilitating its experimental implementation. The improved performance of a Zeeman decelerator in this operating mode is evaluated with the aid of numerical particle trajectory simulations.

\section{\label{sec:model} Phase-space model}

\begin{figure}
\includegraphics{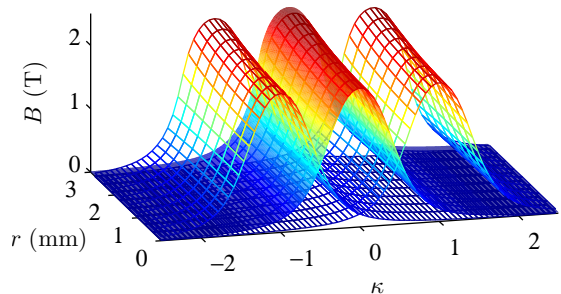}
\caption{\label{fig:Bfield} (Color online) Three-dimensional magnetic field of a solenoid coil for Zeeman deceleration at a current of 300~A (shaded surface). Mesh plots indicate the magnetic fields of neighboring coils. The magnetic fields of the individual coils are not added up vectorially. The coil specifications and positions are the same as in Dulitz et al. \cite{Dulitz2014}.}
\end{figure}

Phase stability ensures that particles within a certain range of relative positions and velocities with respect to a so-called synchronous particle remain together during the successive switching of several deceleration coils. The synchronous particle is an imaginary on-axis particle which experiences precisely those magnetic fields that are calculated for a pulse sequence to achieve a given amount of deceleration or acceleration. In order to achieve phase stability, the solenoid magnetic fields in a multistage Zeeman decelerator are switched in a periodic manner, so that the synchronous particle always moves exactly one coil distance, $d$, before the active coil is turned off. The particle position relative to the center of a deceleration coil can be described by a dimensionless parameter $\kappa=\left(z-z_0\right)/d$ which is related to the particle position on the beam axis, $z$, the center of the active coil, $z_0$, and the center-to-center coil distance, $d$. The change in kinetic energy is determined by the position of the synchronous particle at the switch-off time, $\kappa_0$. 
For infinitely short rise and fall times, phase-stable deceleration is achieved when $\kappa_0$ is chosen such that each solenoid coil is switched off before the synchronous particle reaches the coil center ($\kappa_0$~$<$~0 , see Figure \ref{fig:Bfield}). In this case, more kinetic energy will be removed from particles that are further ahead in the decelerator, while particles lagging behind the synchronous particle will be decelerated less, as they experience a lower average magnetic field gradient. If the positions and velocities of these `non-synchronous' particles are within a certain range relative to the synchronous particle, an oscillatory motion results which is maintained throughout the deceleration process. Wiederkehr et al. \cite{Wiederkehr2010} have devised a one-dimensional model for phase stability that reliably captures this oscillatory behavior.

An accurate description of phase stability becomes more complex if the finite switching times of the current pulses are taken into account. Based on the mean acceleration for each deceleration period, we present an approach to calculate phase stability in a Zeeman decelerator that can cope with arbitrary current pulse shapes. Besides predictions for the longitudinal and the transverse phase stabilities, our model gives an estimate of the overall, six-dimensional phase-space acceptance for each switch-off position $\kappa_0$. This allows for an evaluation of Zeeman-deceleration sequences within a wide parameter range that is complementary to trajectory simulations.

\subsection{\label{sec:1D} Longitudinal and transverse phase-space acceptance}

Our approach builds up on a number of ideas that were previously used to describe the longitudinal and transverse motion in a Stark decelerator \cite{Bethlem2002, van2006a}. Assuming that the longitudinal and the transverse dynamics take place on very different time scales, the longitudinal and the transverse motion can be treated as independent entities. Longitudinal and transverse accelerations, $\bar{a}_z$ and $\bar{a}_r$, are obtained by numerical integration over one period in time, $T$, which is equivalent to the track of the synchronous particle across one coil distance, $d$, at constant particle velocity:

\begin{align}
\bar{a}_i\left(\kappa, r, v_z\right) = \frac{1}{T} \int^{T}_{0} a_i\left(z\left(v_z,t\right),r,t\right) \textrm{d}t
\notag\\ 
\approx \frac{1}{d} \int^{\kappa d + z_0}_{\left(\kappa-1\right)d + z_0} a_i\left(z\left(v_z,t\right),r,t\right) \textrm{d}z 
\label{eq:a}
\end{align}

where $i = z,r$. It is assumed that the longitudinal and transverse particle velocities, $v_z$ and $v_r$, and the transverse positions, $r$, do not change during this interval. Such an approximation is valid for particles with a small magnetic-moment-to-mass ratio and/or for sufficiently high particle velocities. The transverse positions and velocities can be treated as constant for one deceleration period, since off-axis velocities are typically small ($\leq$~20~m/s) and the transverse magnetic field gradients are significantly less than those in the longitudinal direction (see Figure \ref{fig:Bfield}). Changes in the mean longitudinal acceleration as a function of $r$ are less than 20~\% (between $r$~=~0~mm and $r$~=~3~mm) for the coil dimensions shown in Figure \ref{fig:Bfield}.

All calculations in this article were carried out for nitrogen atoms in the metastable $^2$D$_{5/2}$, $M_J$~=~5/2 state. The Zeeman shift, $\Delta E_\textrm{Z}$, for this state \cite{Radford1968} can be linearly approximated by $\Delta E_\textrm{Z} = M_J g_J \mu_\textrm{B} B$, where $M_J = 5/2$ is the projection of the total angular momentum $J$ onto the local magnetic field axis, $g_J$~=~1.20 \cite{BeltranLopez1989} is the Land\'{e} factor, $\mu_\textrm{B}$ denotes the Bohr magneton and $B$ is the magnetic field. Although the results presented in this article are valid for this specific quantum state only, they can be scaled to any other atom or molecule with a linear Zeeman shift. We chose metastable nitrogen due to its relatively small magnetic-moment-to-mass ratio, so that the validity conditions of our model are met. N($^2$D$_{5/2}$) is also a promising candidate for Zeeman deceleration that has not been tackled so far, and we are currently working on the Zeeman deceleration of this species in our laboratory.

\begin{figure}
\includegraphics{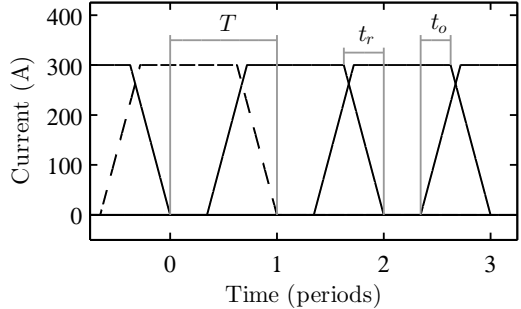}
\caption{\label{fig:period} Temporal profiles of the current pulses for five consecutive deceleration coils as used in the model. The pulse shape for one coil is shown with a dashed line for clarity. Time is scaled to the period $T$, i.e., the time required for the passage of one coil distance. The rise and fall times are denoted with $t_\textrm{r}$, and the time overlap between adjacent pulses, defined as shown in the figure, is referred to as $t_\textrm{o}$.}
\end{figure}

In this article, distances and coil dimensions from our experiment are used \cite{Dulitz2014}, e.g., $d$~=~10.7~mm. Figure \ref{fig:period} illustrates the temporal characteristics of the current pulses. The rise and fall times (assuming equal values in both cases), $t_\textrm{r}$, and the time overlap, $t_\textrm{o}$, are shown on a scale that is related to the time period $T$. In this article, we use $t_\textrm{r}$~=~8~$\mu$s, $t_\textrm{o}$~=~6~$\mu$s and a current of 300~A for each coil, resulting in a maximum mean longitudinal acceleration, $\bar{a}_{z,m}$, of 2.3$\cdot$10$^5$~m/s$^2$ on the beam axis. Mutual inductance effects, which induce additional cusps in the current profiles, were not taken into account for reasons of simplicity. However, in principle, the numerical integration allows for the implementation of any arbitrary time profile, e.g., experimental waveforms.

The cylindrical symmetry of solenoid magnetic fields reduces the complexity of phase-space calculations, since the angular momentum of the particles about the molecular beam axis is conserved, so that the dimensionality of the problem is reduced by two. To obtain information on longitudinal phase stability, the relative mean accelerations 

\begin{equation}
\Delta\bar{a}_z\left(\Delta\kappa,0,v_z\right)=\bar{a}_z\left(\kappa,0,v_z\right)-\bar{a}_z\left(\kappa_0,0,v_z\right), 
\label{eq:daz}
\end{equation}

where $\Delta\kappa = \kappa-\kappa_0$, at $r$~=~0 and $v_z$~=~const. are used to calculate the relative longitudinal positions and velocities of `non-synchronous' particles with respect to the synchronous particle. Trajectories of non-synchronous particles are determined for -4~$<=$~$\Delta\kappa$~$<=$~4 using numerical integration. The separatrix is then given by largest stable orbit around the synchronous particle within this $\Delta\kappa$-interval. Trajectories in the transverse direction at a given longitudinal velocity, $v_z$, are obtained in a similar manner for each switch-off position $\kappa_0$:

\begin{equation}
\Delta\bar{a}_r\left(\kappa_0,\Delta r,v_z\right)=\bar{a}_r\left(\kappa_0,r,v_z\right)-\bar{a}_r\left(\kappa_0,0,v_z\right). 
\label{eq:dar}
\end{equation}

Due to the cylindrical symmetry, the transverse forces on the beam axis ($r$~=~0) vanish, and hence $\bar{a}_r\left(\kappa_0,0,v_z\right)$~=~0.

\begin{figure}
\includegraphics{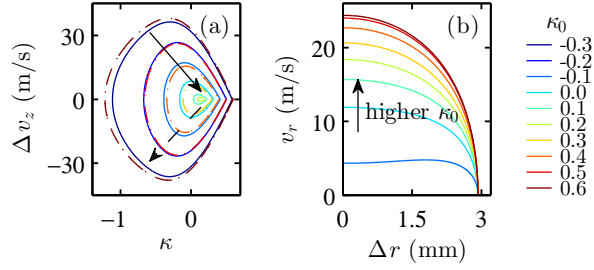}
\caption{\label{fig:separatrices} (Color online) Separatrices in (a) longitudinal and (b) transverse phase space for the Zeeman deceleration of nitrogen atoms in the $^2$D$_{5/2}$, $M_J$~=~5/2 state at different $\kappa_0$ and at a longitudinal velocity of 500 m/s. Particles inside the separatrix revolve in stable orbits around the synchronous particle. Transverse trajectories at $\kappa_0$~$<$~-0.1 are not phase stable. The volume encompassed by the longitudinal separatrices decreases as $\kappa_0$ is increased from -0.3 to 0.2 (solid arrow). There are longitudinal separatrices whose extent increases for $\kappa_0$~$>$~0.1 (dashed lines and dashed arrow in (a)), because the deceleration pulse sequence, applied to achieve a given final velocity, addresses particles in the same phase space as for a lower value of $\kappa_0$, e.g., an identical separatrix centered at $\kappa$~=~-0.1 exists both for $\kappa_0$~=~-0.1 and $\kappa_0$~=~0.4.}
\end{figure}

\begin{figure}
\includegraphics{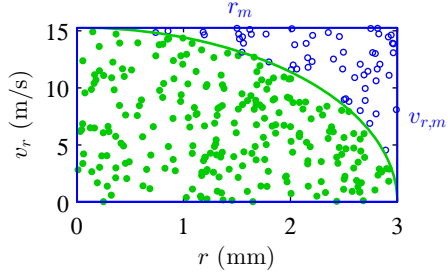}
\caption{\label{fig:MCintegration} (Color online) A schematic illustration of the Monte Carlo numerical integration algorithm used for the calculation of the four-dimensional transverse acceptance. The separatrix (curved green line) is enclosed by a rectangular box (in blue color). The side lengths of the rectangle, $r_m$ and $v_{r,m}$, are determined by the point of origin and the amplitudes of the transverse position and the transverse velocity, respectively.}
\end{figure}

Separatrices, as in Figure \ref{fig:separatrices}, mark the boundaries between stable and unstable trajectories around the synchronous particle. Only trajectories inside a separatrix are phase stable. In Figure \ref{fig:separatrices}, opposite trends are seen for the longitudinal and the transverse stability as $\kappa_0$ is increased. The longitudinal acceptance decreases as $\kappa_0$ is increased from -0.3 to 0.1, i.e., the more kinetic energy is removed during a deceleration step. 
Figure \ref{fig:separatrices} (a) also shows an increase in the longitudinal acceptance for $\kappa_0$~$>$~0.1. In this case, the deceleration pulse sequence, applied to achieve a given amount of deceleration, is effectively the same as for a bunch of particles that revolve around a synchronous particle located at a lower $\kappa$. For example, at $\kappa_0$~=~0.4, a separatrix exists which is centered at $\kappa$~=~-0.1. The change in transverse acceptance can be explained through a change in the shape of the magnetic field (see Figure \ref{fig:Bfield}) from defocusing outside the coil to focusing inside the coil. Hence, the transverse motion is unstable at more negative $\kappa_0$, but increasingly stable the further a particle moves into a coil during one period. The maximum transverse position relative to the beam axis is determined by the distance to the coil walls ($r$~=~3~mm).

Quantitative information on phase-space acceptance can be drawn from the volume that is enclosed within each separatrix. The two-dimensional longitudinal acceptance for each parameter set ($\kappa_0$, $v_z$) is obtained by trapezoidal numerical integration of the points that form the separatrix in longitudinal phase space (see Figure \ref{fig:separatrices} (a)). The transverse phase-space volume, $V_r$, is given by

\begin{equation}
V_r = \int \int \int \int r v_r \textrm{d}r \textrm{d}\theta \textrm{d}v_r \textrm{d}\theta_v.
\end{equation}

The angles $\theta$ and $\theta_v$ are integrated from 0 to 2$\pi$ owing to the cylindrical symmetries of the sought four-dimensional volume. The integrals over $r$ and $v_r$ are taken over the two-dimensional transverse stability region shown within the separatrix in Figure \ref{fig:MCintegration}. 

$V_r$ is evaluated numerically in a Monte-Carlo approach. For this, a uniform random distribution of points $p={r_p,v_{r,p}}$ (all points in Figure \ref{fig:MCintegration}) is drawn within the two-dimensional volume $\mathcal{N}$ given by the distances from the point of origin to the maximum values for the transverse position, $r_m$, and the transverse velocity, $v_{r,m}$. The volume $\mathcal{N}$ entirely encompasses the two-dimensional transverse stability region $\mathcal{M}$ (filled green dots in Figure \ref{fig:MCintegration}). The sought evaluation of $V_r$ therefore reads

\begin{equation}
V_r \approx
(\pi r_m v_{r,m})^2
\frac
{4\pi^2 \sum_{p\in\mathcal{M}} r_p v_{r,p}}
{4\pi^2 \sum_{p\in\mathcal{N}} r_p v_{r,p}}.
\end{equation}

In the calculation, each point $p$ is given a weight $r_p v_{r,p}$ to account for the elementary volume in cylindrical coordinates, while the normalization pre-factor originates from the analytically known volume $\mathcal{N}$.

%

\begin{figure}
\includegraphics{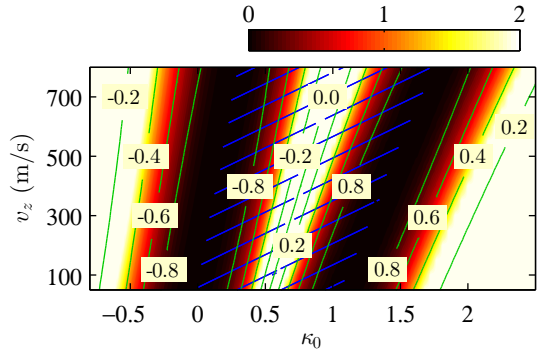}
\caption{\label{fig:acceptz} (Color online) Density plot: longitudinal phase-space acceptance (in m$^2$/s) for Zeeman deceleration/acceleration of N($^2$D$_{5/2}$, $M_J$~=~5/2). Green contour lines (near-vertical lines labeled with white boxes): normalized mean longitudinal acceleration $\bar{a}_z/\bar{a}_{z,m}$ along the beam axis, where $\bar{a}_{z,m}$ is the maximum mean longitudinal acceleration. White boxes give values of $\bar{a}_z/\bar{a}_{z,m}$ for selected contour curves. In the calculation, a cut-off was implemented for acceptances $>$~2~m$^2$/s, as stable longitudinal orbits can extend beyond -4~$<=$~$\Delta\kappa$~$<=$~4. Blue hatches (near-horizontal lines in central region of $\kappa_0$)  mark regions in which longitudinal phase stability is observed  only because the deceleration/acceleration pulse pattern addresses the same particles as in another phase-stable region (no hatches) with the same $\bar{a}_z$, and thus effectively corresponds to a different adaptive $\kappa_0$.}
\end{figure}

\begin{figure}
\includegraphics{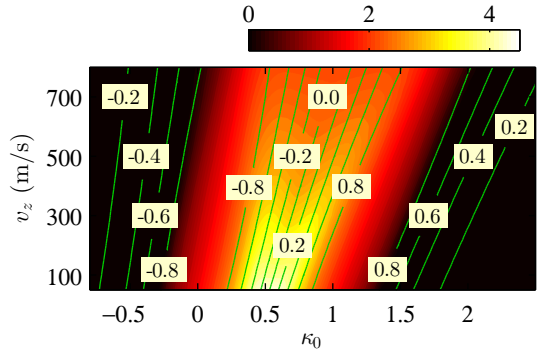}
\caption{\label{fig:acceptx} (Color online) Density plot: transverse phase-space acceptance (in 10$^{-2}$ m$^4$/s$^2$) for Zeeman deceleration/acceleration of N($^2$D$_{5/2}$, $M_J$~=~5/2). As in Figure, \ref{fig:acceptz}, the normalized mean longitudinal acceleration $\bar{a}_z/\bar{a}_{z,m}$ along the beam axis is marked with green contour lines (near-vertical lines labeled with white boxes).}
\end{figure}

Figures \ref{fig:acceptz} and \ref{fig:acceptx} show maps of the longitudinal and transverse acceptances for N($^2$D$_{5/2}$, $M_J$~=~5/2) as a function of the beam velocity, $v_z$, and the switch-off positions of the synchronous particle, $\kappa_0$, respectively. The green (near-vertical) lines indicate the contours of constant mean longitudinal acceleration, $\bar{a}_z$, on the beam axis. For each beam velocity, the variation of the acceptance as a function of $\kappa_0$ is in accordance with the change in the area covered by the separatrices in Figure \ref{fig:separatrices}. However, as the beam velocity decreases, the acceptance profile is shifted to lower values of $\kappa_0$. Both for the longitudinal and the transverse acceptance, this displacement is approximately parallel with the lines of constant longitudinal acceleration. Hence, operating the decelerator at constant mean longitudinal acceleration will ensure that the phase-space acceptance remains nearly unchanged as the beam velocity decreases. 

The velocity dependence of the mean longitudinal acceleration is due to an explicit time dependence of the solenoid magnetic fields caused by the finite rise and fall times for the switching. In the deceleration pulse sequence, the time period, $T$, is increased as the beam velocity decreases to ensure that the synchronous particle always travels one coil distance at each deceleration step. For current pulses such as in Figure \ref{fig:period} with $t_\textrm{r}$~=~8~$\mu$s, the rise and fall times account for a significant fraction of the time period at high velocities causing reduced acceleration. For example, while $T$~=~54 $\mu$s~$\approx 7 t_\textrm{r}$ at 200~m/s, $T$ is as short as 15 $\mu$s~$\approx 2 t_\textrm{r}$ at 700~m/s. Thus, if $\kappa_0$ is kept constant, a synchronous particle experiences much less acceleration at high velocities than at low velocities. At very low beam velocities, the average amount of deceleration approaches the value that is expected for vanishing rise and fall times. While the longitudinal phase-space acceptance does not markedly change as a function of beam velocity (Figure \ref{fig:acceptz}) along a line of constant acceleration, there is a significant decrease in the transverse acceptance (Figure \ref{fig:acceptx}) towards higher velocities. This change is a consequence of the fall times which reduce the transverse particle focusing inside a coil during each deceleration period.

The influence of the beam velocity can be eliminated by linking the experimental rise and fall times to the change in the period, for example, by successively lowering the voltage for the kick interval \cite{Wiederkehr2011} as the particle velocity is decreased. Alternatively, deceleration sequences can be calculated such that the switch-off position $\kappa_0$ follows the change in $\bar{a}_z$ as a function of velocity. The practicality of this scheme is demonstrated in the trajectory simulations below. In addition to that, the use of $\bar{a}_z$ is superior to $\kappa_0$ in the evaluation of alternative Zeeman deceleration schemes (section \ref{sec:applications}), as it provides a direct means of comparison between different operating modes.

In Zeeman deceleration experiments, where the switch-off position inside a coil is defined by a phase angle $\phi_0 = \pi\left(\kappa_0+1/2-v t_\textrm{r}/d\right)$, the beginning and the end of each period in Figure \ref{fig:period} is shifted by $-t_\textrm{r}$, i.e., the end of each period is defined as the time at which the current to the coil is switched off \cite{Hogan2007}. Due to the time overlap between adjacent current pulses, the current for the active coil is then almost constant throughout the interval $T$, and the effect of the rise and fall times from adjacent coils is less. However, the velocity dependence of the phase-space acceptance remains significant irrespective of the definition of the switch-off position, especially at shorter time overlaps and for advanced deceleration schemes (see section \ref{sec:applications}).

\subsection{\label{sec:3D} Overall phase-space acceptance}

The overall phase-space acceptance is \textit{a priori} a complex volume in a six-dimensional phase space, and cannot be correctly evaluated by the product of the longitudinal and transverse acceptances, as shown in Figures \ref{fig:acceptz} and \ref{fig:acceptx}. Instead, in order to evaluate the overall phase-space
acceptance, assumptions on the dynamics of the three-dimensional particle motion are required. We assume that the time for one revolution in transverse phase space, $\tau_r$, is much longer than the time needed for one orbit in longitudinal phase space, $\tau_z$ (adiabatic approximation). In this case, we can calculate the trajectories of non-synchronous particles in longitudinal phase space, and multiply the phase-space volume covered by two adjacent trajectories with the average transverse acceptance that the particles experience during one revolution in longitudinal phase space. The overall six-dimensional phase-space acceptance for each value of $v_z$ and $\kappa_0$ is then obtained by summation over all these sub-volumes.

Our model is compared with numerical three-dimensional particle trajectory simulations \cite{Dulitz2014} for a total of 137 deceleration coils. For the computation of a deceleration pulse sequence, we choose an iso-contour line of $\bar{a}_z$ according to the desired amount of deceleration or acceleration, and vary the switch-off position $\kappa_0$ as a function of the beam velocity (to follow the contour line of constant $\bar{a}_z$). To circumvent deviations from periodicity, the 3~D trajectory simulation starts after the switch-off for the first coil, and it stops when the current of the second to last coil has decayed to zero. An initial velocity of $v_z$~=~800~m/s is chosen for deceleration sequences, such that the final velocity for the maximum amount of deceleration is 200~m/s. Likewise, an initial velocity of $v_z$~=~200~m/s is used for acceleration pulse sequences. Due to the larger beam divergence at low velocities, the program iterates over 15 million metastable nitrogen atoms in the $^2$D$_{5/2}$, $M_J$~=~5/2 state for deceleration, and 105 million particles for acceleration. For a quantitative analysis, we choose uniform random distributions for both the initial particle positions and their velocities in a range that is larger than the maximum extent of the longitudinal and transverse separatrices (Figure \ref{fig:separatrices}), i.e., $\left|\Delta \kappa\right|$~$\leq$~2.3, $\left|\Delta v_z\right|$~$\leq$~50~m/s, $r$~$\leq$~3~mm and $\left|v_r\right|$~$\leq$~25~m/s. 

\begin{figure*} 
\includegraphics{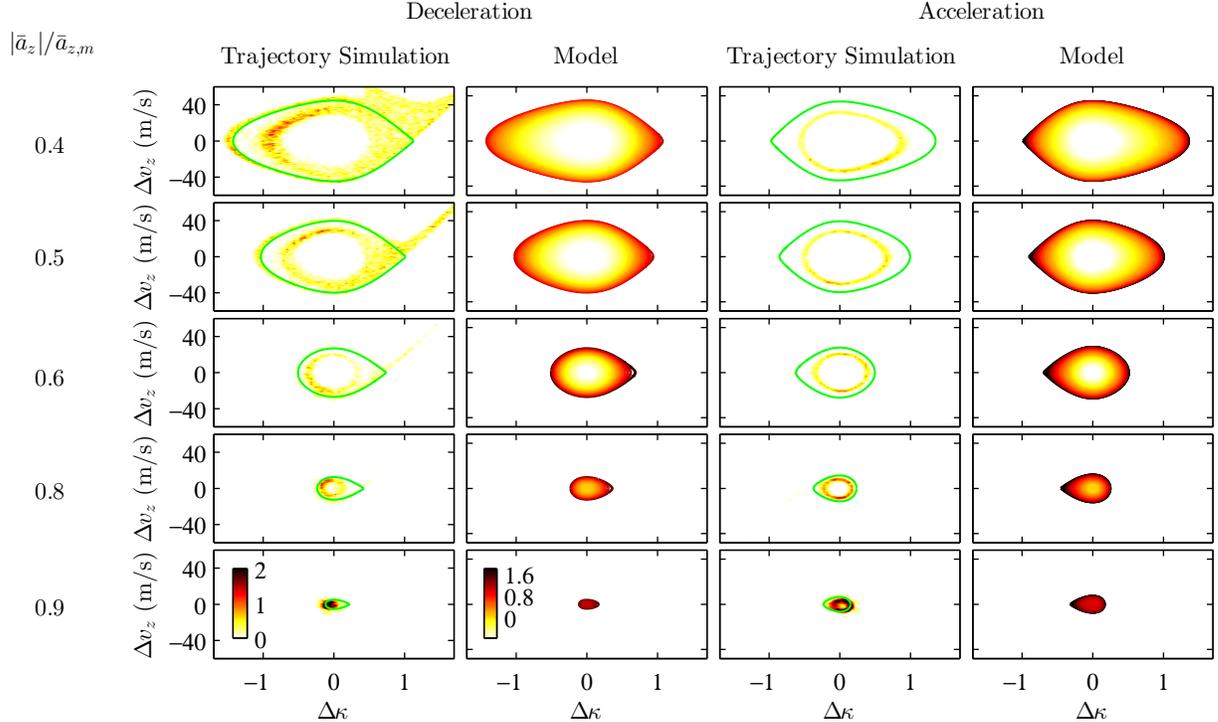}
\caption{\label{fig:fishcomp} (Color online) Longitudinal phase-space distributions of N($^2$D$_{5/2}$, $M_J$~=~5/2) atoms inside the decelerator at different mean longitudinal accelerations, $\bar{a}_z$, that result from 3D trajectory simulations and from the phase-space model. To account for the different number of initial particles in the trajectory simulations, the results for deceleration and acceleration are normalized to the number of particles in the phase-space window at $\bar{a}_z/\bar{a}_{z,m}$~=~-0.6 and 0.6, respectively. Under these conditions, the number of unstable particles remaining in the phase-stable region is expected to be small (see text). The separatrices in longitudinal phase space, as obtained from the model, are shown in the results from trajectory simulations (green (or light-gray) solid-line curves) for comparison. The color scales are referenced to the (scaled) number of particles from the simulation and the transverse acceptance (in 10$^{-2}$ m$^4$/s$^2$) from the model, respectively.}
\end{figure*}

Figure \ref{fig:fishcomp} contrasts longitudinal phase-space distributions obtained from the model and from 3D trajectory simulations. As detailed before, the longitudinal phase-space volume decreases as the mean longitudinal acceleration, $\bar{a}_z$, is increased. The trajectory simulations suggest that the longitudinal separatrices are not uniformly filled for all but the largest amount of deceleration and acceleration ($\left|\bar{a}_z\right|/\bar{a}_{z,m}$~=~0.9 in the figure). Instead, there are two unfilled regions in the longitudinal phase-space diagram, located close to the synchronous particle and near the separatrix. Similar effects were seen by van de Meerakker et al. for a Stark decelerator \cite{van2006a} and by Wiederkehr et al. for a Zeeman decelerator of variable length (56~-~80 deceleration stages) \cite{Wiederkehr2010}. Our model confirms their interpretation of the unfilled region close to the synchronous particle, since comparison with the model input parameters implies that the convex shape of the magnetic fields outside of a solenoid coil (Figure \ref{fig:Bfield}) leads to a net transverse defocusing of low-field-seeking particles over the course of one longitudinal revolution in phase space. Non-synchronous particles at large distances relative to the synchronous particle, $\Delta \kappa$, move further into a coil during a longitudinal orbit which compensates for the transverse defocusing and thus explains the ring of phase-stable trajectories in Figure \ref{fig:fishcomp}. 

\begin{figure}
\includegraphics{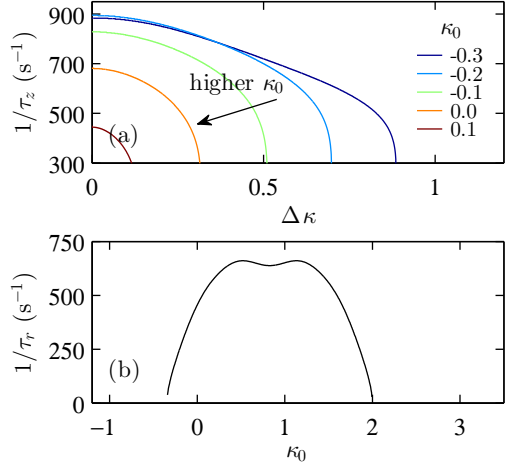}
\caption{\label{fig:tau} (Color online) (a) Inverse time needed for one orbit in longitudinal phase space versus the maximum relative position, $\Delta \kappa$, of non-synchronous particles at different switch-off positions, $\kappa_0$. (b) Inverse time for one revolution in transverse phase space as a function of $\kappa_0$ for a particle moving close to the beam axis ($\Delta r \rightarrow 0$ mm). The data are calculated for the Zeeman deceleration of nitrogen atoms in the $^2$D$_{5/2}$, $M_J$~=~5/2 state at $v_z$ = 500 m/s.}
\end{figure}

As our trajectory simulations mimic the motion through a much longer Zeeman decelerator, the empty phase-space region close to the separatrix (`halo') is much more prominent than in Wiederkehr et al. \cite{Wiederkehr2010}. Following the explanations given for a Stark decelerator \cite{van2006a}, it was assumed that resonant coupling processes lead to unstable trajectories in transverse phase space, and thus induce particle loss during the deceleration process. We also believe that the unfilled phase-space regions in Figure \ref{fig:fishcomp} are due to such resonant couplings which can occur whenever the times for a longitudinal and a transverse revolution in phase space, $\tau_r$ and $\tau_z$, become very similar. From Figure \ref{fig:tau}, we see that $\tau_r$ and $\tau_z$ can indeed become comparable under certain conditions for $\Delta \kappa$ and $\kappa_0$. Our model is unable to capture these effects, because it explicitly assumes adiabatic behavior, i.e., $\tau_r \gg \tau_z$ (see above). We believe that the unfilled phase-space regions close to the separatrix in Figure \ref{fig:fishcomp} can be explained through such a mechanism. 

\begin{figure}
\includegraphics{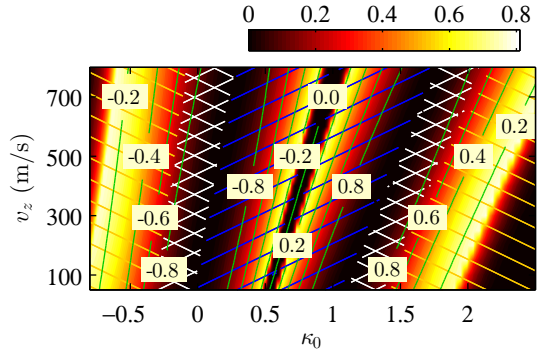}
\caption{\label{fig:overallacceptance} (Color online) Density plot: overall phase-space acceptance (in 10$^{-2}$~m$^6$/s$^3$) for Zeeman deceleration/acceleration of N($^2$D$_{5/2}$, $M_J$~=~5/2). As in Figures \ref{fig:acceptz} and \ref{fig:acceptx}, the normalized mean longitudinal acceleration $\bar{a}_z/\bar{a}_{z,m}$ along the beam axis is marked with green contour lines ((near-vertical lines labeled with white boxes). Orange hatches (light near-horizontal lines at left- and right-hand side of figure): regions in which only particles with very high longitudinal velocities or large displacements with respect to the synchronous particle are phase stable. White crossed hatches: particle motion close to the synchronous particle is phase stable. Blue  hatches (dark, near-horizontal in central region of figure): regions in which phase stability is observed owing to a deceleration/acceleration pulse sequence which addresses particles in the same phase space as for the same value of $\bar{a}_z$ in a region marked with orange or white crossed hatches, respectively.}
\end{figure}

Figure \ref{fig:overallacceptance} shows the overall phase-space acceptance that is obtained from the model at different beam velocities, $v_z$, using the same conditions as above. In agreement with Figures \ref{fig:acceptz} and \ref{fig:acceptx}, there is an explicit velocity dependency of the overall acceptance in terms of $\bar{a}_z$. However, following any contour line of constant mean longitudinal acceleration, the overall acceptance does not markedly change as a function of beam velocity. 

The phase-space acceptance highlighted with blue hatches (dark lines, near horizontal) in the central region of Figure \ref{fig:overallacceptance} is a reflection of the phase stability for the same $\bar{a}_z$ that is already observed in the regions highlighted with white cross-hatching and orange hatching, since the applied deceleration pulse sequence effectively addresses particles in the same phase-space volume. Neglecting this `fake' additional phase-space acceptance, there is only a limited parameter range for Zeeman deceleration and acceleration, where the overall six-dimensional phase-space acceptance is non-zero. This is in accordance with the partial overlap between the regions of maximum longitudinal and transverse acceptance (cf. Figures \ref{fig:acceptz} and \ref{fig:acceptx}). Phase stability close to the synchronous particle is seen in an even smaller region (white crossed hatches in Figure \ref{fig:overallacceptance}).

For experiments in which the transmitted beam is used to study collisions, these results imply that Zeeman deceleration and acceleration should ideally be carried out at 0.8 $\leq\left|\bar{a}_z\right|/\bar{a}_{z,m}\leq$~1.0 (white cross-hatched regions in Figure \ref{fig:overallacceptance}) to obtain a superior kinetic energy resolution while maintaining an almost uniform distribution in phase space. If maximum transmission is the major goal, e.g., for trapping experiments, a lower mean longitudinal acceleration would be more advantageous (orange-hatched regions in Figure \ref{fig:overallacceptance}). The maximum overall acceptance is obtained at a mean longitudinal acceleration of $\left|\bar{a}_z\right|/\bar{a}_{z,m}$=~0.3, but it comes at the expense of a large spread in relative particle positions and velocities with respect to the synchronous particle. 

\begin{figure}
\includegraphics{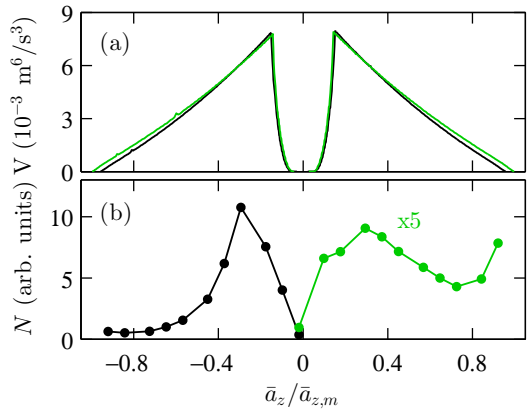}
\caption{\label{fig:quantcomp} (Color online) (a) Overall phase-space acceptance, $V$, obtained from the phase-space model and (b) the number of transmitted particles, $N$, in the trajectory simulation (dots) versus the normalized mean longitudinal acceleration, $\bar{a}_z/\bar{a}_{z,m}$. $V$ is proportional to $N$, since the trajectory simulation was carried out with uniform initial position and velocity distributions. Green (or mid-gray) traces and black traces in (a) correspond to the overall acceptance at $v_z$~=~200~m/s and 800~m/s, respectively. The number of decelerated (accelerated) particles in (b) is derived from the number of particles in the phase-space windows shown in Figure \ref{fig:overallacceptance}; the solid lines are a guide to the eye only. The number of accelerated particles (green (or mid-gray) dots) is upscaled for visibility.}
\end{figure}

Considering the limited validity of our model, we do not expect quantitative agreement between the output of the model and results from trajectory simulations.
However, Figure \ref{fig:quantcomp} shows that the general trends for the model and the simulation are very similar, e.g., the position of the maximum in phase-space acceptance in terms of $\bar{a}_z$ and the decrease in acceptance towards larger values of the mean longitudinal acceleration. Despite the limitations of our model, we are confident that it can provide an order-of-magnitude estimate of the overall phase-space acceptance in a Zeeman decelerator.

Again, Figure \ref{fig:quantcomp} (a) illustrates that the six-dimensional phase-space volume does not notably change between $v_z$~=~200~m/s (green (or mid-gray) lines) and 800~m/s (black lines). In this representation, it also becomes obvious that, for any given value of $\left|\bar{a}_z\right|/\bar{a}_{z,m}$, the overall phase-space acceptance is the same for Zeeman deceleration and acceleration. However, in the trajectory simulations (Figures \ref{fig:fishcomp} and \ref{fig:quantcomp}), we see that a lot more particles are transmitted during the deceleration sequence than during acceleration. In the case of Zeeman deceleration, a significant number of particles is captured within the phase-stable region, even though their trajectories are not phase-stable, especially at low $\bar{a}_z$. This effect is also apparent in the distinct tail of particles outside the longitudinal separatrix (Figure \ref{fig:fishcomp}) indicating that particles are leaving the phase-stable region even after having passed more than a hundred deceleration stages. As the initial velocity for acceleration is much lower than for deceleration -- 200~m/s as compared to 800~m/s -- non-synchronous particles, that do not meet the conditions for phase-space acceptance, have much more time to leave the phase-stable region or to hit the walls during the acceleration process.

Various experimental studies have demonstrated that the intensity of the decelerated signal changes according to the amount of deceleration \cite{Hogan2008a, Narevicius2008, Narevicius2008a, Wiederkehr2010, Wiederkehr2011, Dulitz2014}. Due to the different experimental arrangements, a direct comparison with the model results in this article is difficult. Although phase-space acceptance is not pronounced in our short 12-stage Zeeman decelerator in Oxford, 
our experimental results on ground-state hydrogen atoms \cite{Dulitz2014} clearly show a signal decrease by about a factor of two from $\kappa_0$~=~-1 to $\kappa_0$~=~0 which is what would be expected from the model. 

\begin{figure}
\includegraphics{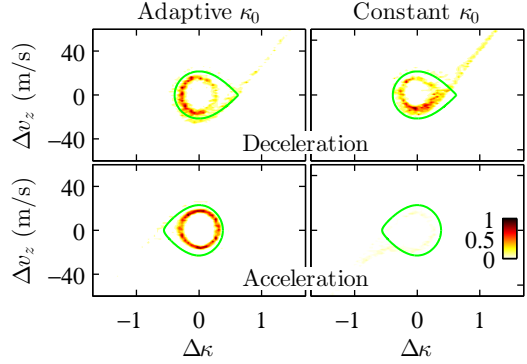}
\caption{\label{fig:constantk} (Color online) Longitudinal phase-space distributions for Zeeman deceleration (upper panel) and acceleration (lower panel) of N($^2$D$_{5/2}$, $M_J$~=~5/2) atoms, respectively. Results are obtained from particle trajectory simulations using an adaptive (left column) and a constant (right column) switch-off position, $\kappa_0$. In the case of an adaptive $\kappa_0$, a mean longitudinal acceleration of $\left|\bar{a}_z\right|/\bar{a}_{z,m}$=~0.7 is used. The constant values for $\kappa_0$ are chosen such that the final velocities are the same in both modes of operation, i.e., $\kappa_0$~=~-0.1  for deceleration (800~m/s~$\rightarrow$~=~410~m/s) and $\kappa_0$~=~1.8 for acceleration (200~m/s$\rightarrow$~710~m/s). Normalization is the same as in Figure \ref{fig:fishcomp}, but the color scale is adjusted to increase the contrast. For comparison, longitudinal separatrices from the model are shown as solid green (or light-gray) curves.}
\end{figure}

The consequences of using a constant switch-off position, $\kappa_0$, i.e., successively increasing $\left|\bar{a}_z\right|$ during a Zeeman deceleration or acceleration sequence, are two-fold. The upper (lower) panel in Figure \ref{fig:constantk} shows results from trajectory simulations in which an adaptive and a constant $\kappa_0$ value are chosen for Zeeman deceleration (acceleration), such that the final velocity is the same. In the case of Zeeman acceleration, we observe a dramatic decrease in the number of phase-stable particles in the constant $\kappa_0$ mode of operation while there is little change in the phase-space distribution for Zeeman deceleration. The behavior shown for both Zeeman deceleration and acceleration at constant $\kappa_0$ can be attributed to a change from a larger to a smaller phase-space volume meaning that particles originally located in a phase-stable region will find themselves in an unstable region at the end of a deceleration or acceleration sequence (cf. Figure \ref{fig:fishcomp}). Likewise, a lot of particles are ejected from the decelerator at the beginning of a deceleration or acceleration sequence although their trajectories would be phase stable at a later point in time. However, in the acceleration process, the more pronounced velocity dependence along with the greater velocity change ($\Delta v_z$~=~-390~m/s for deceleration, $\Delta v_z$~=~510~m/s for acceleration) causes a stronger decrease in phase-space acceptance, and results in less particle transmission through the decelerator. In this specific case, the
results of applying  an adaptive $\kappa_0$ (a constant $\bar{a}_z$) only imply a significant advantage for Zeeman acceleration, and little advantage for deceleration. However, the use of this procedure  generally simplifies data interpretation in terms of phase-space acceptance, and it helps to find improved deceleration schemes, as demonstrated in Section \ref{sec:applications}.

\subsection{\label{sec:timeoverlap} Influence of the time overlap on the overall phase-space acceptance}

\begin{figure}
\includegraphics{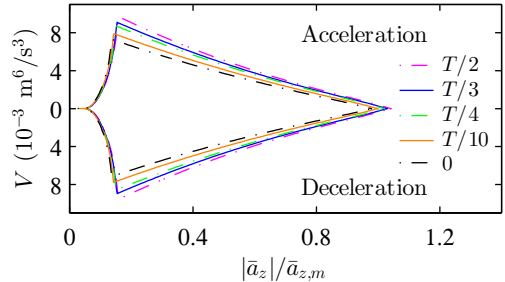}
\caption{\label{fig:0r} (Color online) Overall phase-space acceptance, $V$, as a function of normalized mean longitudinal acceleration, $\left|\bar{a}_z\right|/\bar{a}_{z,m}$, for Zeeman deceleration (bottom) and acceleration (top) of N($^2$D$_{5/2}$, $M_J$~=~5/2) atoms at different overlap times between adjacent coils, as indicated in the legend. Overlap times are linked to the time period, $T$. Infinitely fast rise and fall times are assumed.}
\end{figure}

Thus far, we have neglected the effect of the time overlap between the current pulses of adjacent coils, $t_o$, and used a constant value of $t_o$~=~6~$\mu$s. However, following the same arguments as for the rise and fall times, a constant time overlap will have a greater impact at higher beam velocities, and its influence will be negligible at low velocities. As the particles are typically quite far away from the center of the neighboring coil at the time it is switched on, we do not expect to see as strong a velocity dependence of $\bar{a}_z$ as for the rise and fall times, $t_r$. The influence of the time overlap on the overall phase-space acceptance is shown in Figure \ref{fig:0r} for $t_r$~=~0. In this idealized case, the acceptance only depends on changes in the time overlap. Furthermore, a coupling of the time overlap to the period ($t_o \propto T$) eliminates the velocity dependence of $\bar{a}_z$. Figure \ref{fig:0r} illustrates that the overall acceptance steadily increases as the time overlap becomes a larger fraction of the time period. The increase is small (about 20~\% from $t_o$~=~0 to $t_o$~=~$T/2$) and the maximum phase-space acceptance is shifted to higher mean longitudinal accelerations, equally for Zeeman deceleration and acceleration. We observe a similar effect when both the switching times and the time overlap are linked to the time period, e.g., $t_r$~=~$t_o$~=~$T/2$. In this case, however, the overall acceptance is lower due to the contribution of the rise and fall times. With regard to Figures \ref{fig:overallacceptance} and \ref{fig:quantcomp}, the time overlap ensures that the magnitude of the overall acceptance is nearly constant at all beam velocities. Without this contribution, the rise and fall times cause a gradual decrease in acceptance towards higher velocities (not shown). The increase in overall acceptance as a function of the time overlap is mainly a longitudinal effect. In fact, as $t_r$ is increased, the transverse acceptance decreases due to the larger contribution from transversely defocusing magnetic fields outside of a coil. In contrast to that, the additional magnetic field from the adjacent coil raises the potential hill in the longitudinal direction, so that particles with a position further ahead in the decelerator will still be captured in longitudinal phase space. As the contribution is small and the heating of solenoid coils during operation is generally a major concern, we do not think that an excessive increase in overlap times will contribute much to the success of future deceleration experiments. However, this effect should be kept in mind when considering alternative switching schemes. In addition to that, a time overlap between neighboring coils also guarantees a residual magnetic field to prevent Majorana spin-flip transitions \cite{Vanhaecke2007a, Hogan2008a}. 

\section{\label{sec:applications} Applications of the model to advanced deceleration schemes}

In the previous section of this article, we have demonstrated that our model can provide a better understanding of phase stability in a Zeeman decelerator. We saw that there are a number of limitations to the normal operating mode, especially the empty regions in phase space and the large variation of the overall acceptance as a function of the mean longitudinal acceleration. Alternative Zeeman deceleration schemes may resolve these problems. In the following, we use our model to evaluate the feasibility of two advanced operating modes that are inspired by schemes developed for Stark deceleration. Thereby, we demonstrate that our model can serve as a valuable tool to assess the performance of different Zeeman deceleration sequences. In section \ref{sec:acceldecel}, we present a switching scheme for Zeeman deceleration that makes more effective use of the phase-space characteristics in the normal deceleration mode, and we show that this approach provides superior phase-space characteristics that are especially well suited for collision experiments.

\subsection{\label{sec:Starkswitching4Zeeman} Mimicking Stark deceleration sequences}

There are a lot of parallels between Stark and Zeeman deceleration, and a number of authors have already compared the two techniques in detail \cite{Wiederkehr2010, Hogan2011, van2012}. Differences arise from technical restrictions, for example, inductance and the resistive heating of solenoid coils in a Zeeman decelerator thereby limiting the switching times and the overall duration of a current pulse, respectively. In addition, the longitudinal and transverse properties of electric and magnetic fields are dissimilar, so that a switching scheme may work reliably in one deceleration technique but not in the other.  

In a Stark decelerator, alternating deceleration stages are electrically connected. Thus, when the electric field stages are successively switched between a high voltage and a low voltage (or ground) configuration, a potential hill is created at both sides of the synchronous particle. In the $s$~=~1 mode of operation, the switching takes place as soon as the synchronous particle has traveled exactly one electrode-to-electrode distance, $d$, while in the $s$~=~3 configuration, the electrodes are switched after the particles have covered a distance of 3$d$, so that the particles have to surmount an additional potential hill during each deceleration step. The latter mode of operation provides additional transverse confinement, and it effectively decouples the longitudinal and the transverse dynamics. The $s$~=~1 scheme enables guiding, i.e., an equal amount of acceleration and deceleration per switching stage ($\bar{a}_z$~=~0), so that the actual beam velocity is not changed during the switching sequence. Guiding can prove useful in the initial characterization of a supersonic beam. However, as shown in Section \ref{sec:model}, switching in the normal Zeeman deceleration mode does not result in phase-stable motion at $\bar{a}_z$~=~0.

\begin{figure}
\includegraphics{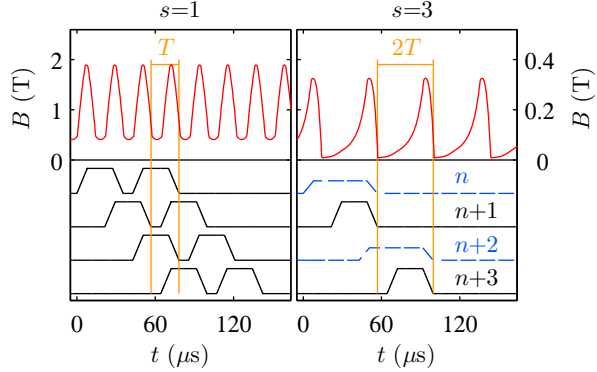}
\caption{\label{fig:Starkseq} (Color online) Zeeman deceleration sequences imitating the $s$~=~1 (left column) and $s$~=~3 (right column) operating modes of a Stark decelerator. The lower traces schematically indicate the switching times of four consecutive deceleration coils, and the lines (red online) in the upper panel illustrate the magnetic field experienced by the synchronous particle (initial velocity of 500~m/s). A guiding sequence ($\bar{a}_z$~=~0) is chosen for the $s$~=~1 mode. In the $s$~=~3 sequence, $\bar{a}_z/\bar{a}_{z,m}$=~-0.14 and a current of 150~A is applied to every focusing coil (dashed curves, blue online). All other coils are operated at 300~A. The horizontal bars indicate the time periods $T$ and $2 T$ that are used to obtain the mean longitudinal acceleration, $\bar{a}_z$.}
\end{figure}

In the following, we will mimic both the $s$~=~1 and the $s$~=~3 mode but, in view of ohmic heating, we will restrict the switching so that each coil is turned on for a maximum duration of two periods, $T$, as compared to the normal mode of operation. Figure \ref{fig:Starkseq} shows schematic pulse sequences for both schemes (lower panel), and it illustrates the corresponding magnetic field experienced by the synchronous particle under these conditions (upper panel). The $s$~=~1 mode is not very different from the normal mode of operation, except that each coil is turned on again after a break of one period, $T$. The duration of the second pulse is synchronized with the other switching times, so that its duration is also of time $T$. In the $s$~=~3 sequence, every second coil is turned on for a duration of two periods (blue dashed curves), while every other coil is switched as usual. In contrast to the $s$~=~3 sequence in a Stark decelerator, every coil represents one potential hill, so that our $s$~=~3 mode is essentially $s$~=~2. To study focusing in the transverse direction, we choose one configuration in which all coils are operated at 300~A, and another in which a current of 150~A is applied to those coils with an extended pulse duration of 2$T$ (dashed curves, `focusing coils'). Similar $s$~=~3 Zeeman deceleration schemes have been described by Wiederkehr et al. \cite{Wiederkehr2010}, but either every second coil was not switched, or the coils were placed at larger distances with respect to each other. From the arguments in section \ref{sec:3D}, it seems reasonable to keep the particles within the active coil for as much time as possible to increase transverse focusing. Therefore, we decided not to change the coil configuration with respect to the normal mode of operation.

\begin{figure}
\includegraphics{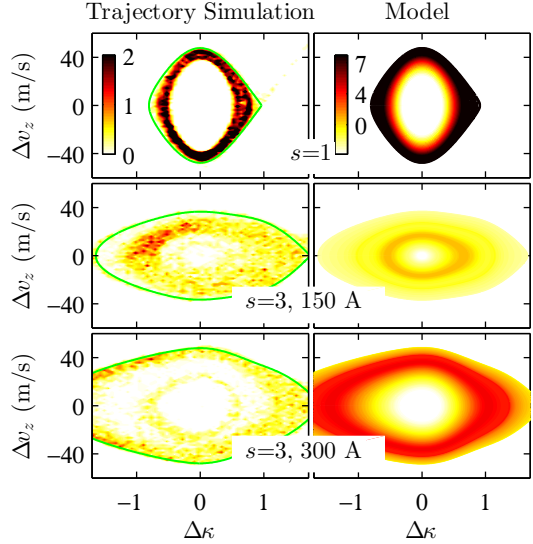}
\caption{\label{fig:Starkfish} (Color online) Longitudinal phase-space distributions for Zeeman guiding and deceleration of N($^2$D$_{5/2}$, $M_J$~=~5/2) atoms in the $s$~=~1 (top) and $s$~=~3 (middle and bottom) modes of operation, respectively. The focusing coils in the middle panel are operated at 150~A. Otherwise, a current of 300~A is applied to the coils. The mean longitudinal acceleration is $\bar{a}_z/\bar{a}_{z,m}$=~0 ($s$~=~1, top), -0.21 ($s$~=~3, 150~A, middle) and -0.23 ($s$~=~3, 300~A, bottom). Trajectory results and longitudinal separatrices from the model (green (or light gray) solid curves) are shown in the left column. The output from the model is displayed in the right column. The color scales are referenced to the number of particles from the simulation and the transverse acceptance (in 10$^{-3}$ m$^4$/s$^2$) from the model, respectively. For the sake of clarity, the color code is matched to the maximum value among the three plots in each column.}
\end{figure}

The implementation of the $s$~=~1 and $s$~=~3 sequences in the model is analogous to what has been described for the normal mode of operation, but care must be taken in the determination of the mean longitudinal acceleration, $\bar{a}_z$. While averaging $\bar{a}_z$ over one period is sufficient in the $s$~=~1 scheme, two periods (2$T$ or 2$d$) must be taken into account in the $s$~=~3 mode, because the pulse patterns for two adjacent coils differ (cf. Figure \ref{fig:Starkseq}). Again, we compare the results of our model with trajectory simulations. In this case, an initial velocity of 500~m/s is chosen for the guiding, deceleration or acceleration of 10 million nitrogen atoms in the $^2$D$_{5/2}$, $M_J$~=~5/2 state through a 137-stage Zeeman decelerator (configuration as in section \ref{sec:3D}). 

Selected longitudinal phase-space distributions obtained from model and trajectory calculations are highlighted in Figure \ref{fig:Starkfish}. There is very good agreement for the extent of the longitudinal phase space, and the empty phase-space region in the center of the phase space is well reproduced by the model. As expected for guiding in the $s$~=~1 mode, the longitudinal phase space is mirror-symmetric with respect to the $v_z$-axis, which indicates that the potential hills both in front and behind the synchronous particle are of similar shape. Due to the vectorial addition of the magnetic fields, the region in between two coils is even more transversely defocusing than in the normal mode of operation thus explaining the large depleted phase space in the center of the distribution. Similar features are also observed for deceleration and acceleration pulse sequences. Unfilled central regions are less of a problem in the $s$~=~3 schemes. As seen for the normal mode of operation, the trajectory data reveal unfilled regions in phase space, e.g., for $s$~=~3, 300 A in Figure \ref{fig:Starkfish}, that differ from the model predictions. Again, we assume that these regions are dominated by motion that cannot be described within the adiabatic approximation of our model. The longitudinal phase space is much more evenly filled when a current of 150~A is applied to the focusing coils (also at other mean longitudinal accelerations that are not shown here). This indicates that the longitudinal and the transverse dynamics may be uncoupled more effectively if the currents used for deceleration and for transverse confinement are not the same.

\begin{figure}
\includegraphics{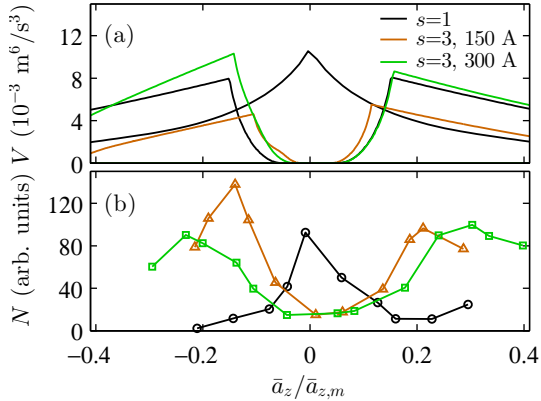}
\caption{\label{fig:Starkaccept} (Color online) (a) Overall phase-space acceptance, $V$, for the $s$~=~1 and $s$~=~3 modes as a function of the normalized mean longitudinal acceleration, $\bar{a}_z/\bar{a}_{z,m}$, calculated using the model for phase stability; see legend for assignments. (b) Number of transmitted particles, $N$, for the same modes of operation (color code as in (a)) versus $\bar{a}_z/\bar{a}_{z,m}$ that is obtained from the trajectory simulation (initial velocity of 500~m/s) by counting the number of particles in the phase-space window (cf. Figure \ref{fig:Starkfish}). The solid lines are a guide to the eye only. The scaling of the trajectory data is not the same as in Figure \ref{fig:quantcomp}. Here, a value of $N$~=~10 corresponds to a transmission of 100 phase-stable particles per 1 million initial particles. The final velocities are within a range of 240~-~690~m/s for the $s$~=~3, 300~A mode.}
\end{figure}

As in the normal mode of operation, we observe a distinct velocity dependence of the phase-space acceptance in all three alternative switching schemes (not shown). Figure \ref{fig:Starkaccept} illustrates that there is a very good correlation between model and trajectory simulation results concerning the overall phase-space acceptance as a function of $\bar{a}_z$. The overall acceptance for the $s$~=~3 schemes is highest at $\left|\bar{a}_z\right|/\bar{a}_{z,m}$=~0.2-0.3, with the 150~A data reaching the maximum at lower mean longitudinal acceleration. For the $s$~=~1 mode, the acceptance peaks at $\bar{a}_z$~=~0, with a significantly higher phase-space acceptance than in the normal mode of operation. Unfortunately, the acceptance decays quickly towards higher mean longitudinal accelerations, especially in the trajectory simulations. As for the normal mode of operation, there are recurring regions of phase-space acceptance for the same value of $\bar{a}_z$ (not shown in Figure \ref{fig:Starkaccept}). In the $s$~=~1 mode, there are other acceptance regions for deceleration and acceleration in addition to that. To access these regions, the pulse sequences are chosen such that the reswitching of each coil has no effect on the particles. The magnetic fields in these sequences then effectively correspond to those in the normal mode of operation thus explaining why the maximum acceptance is the same as in Figure \ref{fig:quantcomp}. In this case, however, the use of an $s$~=~1 mode proves unnecessary.

In general, due to the low overall acceptance beyond $\bar{a}_z$~=~0 and the unfilled regions in phase space, the $s$~=~1 mode does not seem suitable for deceleration or acceleration experiments. However, owing to the very high acceptance at $\bar{a}_z$~=~0, the guiding mode may still be useful in the characterization of a supersonic beam. The $s$~=~3 scheme seems to be promising for Zeeman deceleration at low $\bar{a}_z$, especially for a focusing current of 150~A, because the longitudinal phase space is more uniformly filled. The main disadvantage of this approach is the much smaller maximum amount of deceleration/acceleration that can be attained in comparison to the normal mode of operation. Essentially twice as many coils are needed, since every second coil is used for transverse focusing instead of deceleration.


\subsection{\label{sec:acceldecel} Acceleration-deceleration switching scheme}

In section \ref{sec:model}, we saw that the phase-space acceptance in a Zeeman decelerator varies as a function of the mean longitudinal acceleration, $\bar{a}_z$. Particularly at low $\bar{a}_z$, the extent of the phase-stable region in ($\Delta\kappa,v_z$)-space is huge compared to higher amounts of acceleration; and above all, the 6D phase-space volume is not evenly filled. These phase-space characteristics are not practical for collision experiments, where ideally, the phase-space volume should stay constant as the kinetic energy of the beam is changed. Here, we describe an approach that makes it possible to tune the final velocity of a supersonic beam over a wide range while keeping the phase-space volume nearly constant. This scheme, which is based upon the switching between Zeeman acceleration and deceleration, is also very easy to implement in existing Zeeman decelerator experiments as it only relies on changes in the computed deceleration pulse sequence.

From the analysis of the normal mode of operation, we know that the occupied phase-stable regions for Zeeman deceleration and acceleration are mirror-symmetric (Figure \ref{fig:fishcomp}) if the amount of acceleration is equal to the amount of deceleration ($+\bar{a}_z=-\bar{a}_z$). Equally, for a given $\left|\bar{a}_z\right|$, the overall acceptance in our model is virtually the same for deceleration and acceleration (Figure \ref{fig:quantcomp}). Assuming that our experiment consists of a set number of coils, $m$, we can choose a specific mean longitudinal acceleration, $\left|\bar{a}_z\right|$, and use the first $n$ stages for Zeeman acceleration and the remaining $m-n$ coils for deceleration. Tuning of the final beam velocity is then achieved solely by changing the relative number of coils for acceleration and deceleration, $n$ and $m-n$, respectively. The minimum (maximum) velocity is given by the final velocity that can be attained if all coils are used for deceleration (acceleration). This scheme not only allows for efficient deceleration and acceleration at low or zero effective mean acceleration, $\bar{a}_{z,e}$, but it also makes it possible to stay within the same phase-stable region throughout the acceleration-deceleration sequence. In our trajectory simulations for this scheme, we carry out all acceleration steps before we switch over to the deceleration mode. The two coils at the changeover between the acceleration and deceleration stages are not operated to allow for the synchronous particle to travel one coil distance within a period, $T$. The experimental arrangement for the simulations is the same as in section \ref{sec:3D} (137 stages). In each numerical calculation, 10 million N($^2$D$_{5/2}$, $M_J$~=~5/2) atoms with an initial velocity of 500~m/s are propagated from the switch-off of the first coil until the current of the second to last coil has reached zero.

\begin{figure}
\includegraphics{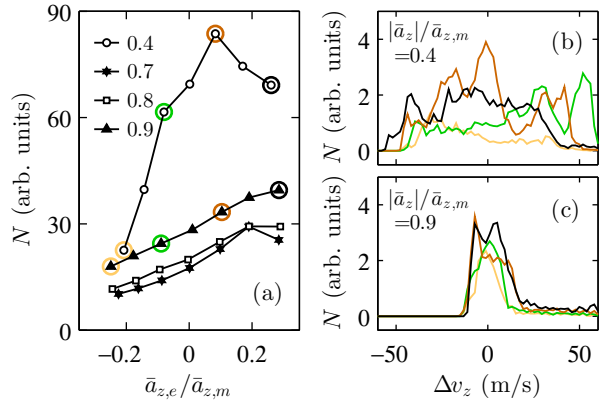}
\caption{\label{fig:accdec} (Color online) (a) Trajectory simulation results for Zeeman deceleration of N($^2$D$_{5/2}$, $M_J$~=~5/2) atoms in the acceleration-deceleration mode using $\left|\bar{a}_z\right|/\bar{a}_{z,m}$=~0.4, 0.7, 0.8 and 0.9 as indicated in the legend. The particle numbers within the phase-space window (cf. Figure \ref{fig:fishcomp}), $N$, are plotted against the normalized effective mean acceleration, $\bar{a}_{z,e}/\bar{a}_{z,m}$, where $\bar{a}_{z,e}/\bar{a}_{z,m}$~=~0 corresponds to a velocity of 500~m/s. The solid lines are a guide to the eye only. The scaling of the trajectory data is the same as in Figure \ref{fig:Starkaccept}. (b) and (c) Velocity distributions, $\Delta v_z$, for $\left|\bar{a}_z\right|/\bar{a}_{z,m}$=~0.4 and 0.9 at different $\bar{a}_{z,e}/\bar{a}_{z,m}$, respectively. The line shadings (colors online) in (b) and (c) correspond to the shading (coloring) of the circles in (a).}
\end{figure}

Figure \ref{fig:accdec} highlights trajectory simulation results obtained for the acceleration-deceleration switching scheme at four different $\left|\bar{a}_z\right|$ values. The relative number of phase-stable trajectories is shown as a function of the effective mean acceleration averaged over the acceleration and deceleration sequences. We can see that this operating mode indeed allows for a wide tuning range in terms of the final beam velocity, including acceleration, deceleration and guiding. For $\left|\bar{a}_z\right|/\bar{a}_{z,m}$=~0.4, the transmitted particle numbers in the phase-stable region are comparable to the $s$~=~1 and $s$~=~3 modes of operation (Figure \ref{fig:accdec} (a)). However, the obtained particle velocity distributions are very broad and not uniform (Figure \ref{fig:accdec} (b)). In contrast to that, switching at a higher mean longitudinal acceleration of $\left|\bar{a}_z\right|/\bar{a}_{z,m}$=~0.9 (Figure \ref{fig:accdec} (c)) yields a very narrow velocity distribution which is centered near zero. In addition to that, the number of transmitted particles for  $\left|\bar{a}_z\right|/\bar{a}_{z,m}$=~0.7~-~0.9 changes only by a maximum factor of two between the minimum and the maximum final velocities considered, which is much smaller than in any other operating mode presented in this article (cf. Figures \ref{fig:quantcomp} and \ref{fig:Starkaccept}). Owing to the small and almost constant velocity spread, switching at $\left|\bar{a}_z\right|/\bar{a}_{z,m}$=~0.9 would clearly be the method of choice for collision experiments.

The output of the simulation can be understood from the corresponding phase-space distributions in the normal mode of operation (see Figure \ref{fig:fishcomp}) which predicts higher particle numbers and broader velocity distributions at low values of $\left|\bar{a}_z\right|$. It is somewhat surprising to see that more particles are transmitted at $\left|\bar{a}_z\right|/\bar{a}_{z,m}$=~0.9 (triangles in Figure \ref{fig:accdec})(a)) than in the intermediate regime (open squares and stars in the figure).
This may be related to the more uniform phase-space distribution at high mean accelerations which helps to match the phase-space distributions for acceleration and deceleration.

\section{\label{sec:conclusion} Conclusion}

In this paper, we have introduced a new formalism to calculate the longitudinal, transverse and overall phase-space acceptance in a Zeeman decelerator. Simulations using this model have enabled us to deduce a number of significant physical insights for particles moving through a sequence of pulsed fields, and these have implications not only for Zeeman decelerators using pulsed magnetic fields, but also Stark decelerators using electric fields. These physical insights and their implications are as follows:

(1) We have simulated the behavior of a bunch of magnetic particles moving through an array of solenoid coils, to which pulsed currents are applied such that the particles experience a series of pulsed magnetic fields leading either to a deceleration or an acceleration of the particles. The field sequence is designed such that a `synchronous' (on-axis) particle has traveled exactly one coil distance in between successive pulses.  Since the particle velocity decreases over the course of a deceleration sequence, it takes longer for the particle to travel one coil distance and, hence, the period must increase as the sequence progresses. In this article, we show that the particles do not experience a constant mean deceleration for each pulse if the applied current pulses have finite rise and fall times. This effect leads to a change in the phase-space-acceptance volume - defined in terms of the range of different longitudinal and transverse positions and 
velocities that have stable trajectories - as the sequence of pulses progresses. The effects of pulses with finite rise and fall times has not been considered in either Zeeman or Stark decelerators in previous work; the model and simulations presented here are able to catch and overcome this difficult problem efficiently, both for deceleration and acceleration sequences. This physical understanding leads to the technical conclusion that a beam with well-defined properties is produced by adjusting the pulse durations in such a way as to maintain constant mean deceleration/acceleration.

(2) Our model shows that the phase-space-acceptance volume is virtually identical for an acceleration sequence compared to a deceleration sequence provided one chooses the same magnitude of the mean acceleration/deceleration. This leads to the proposal, ratified by simulations, that the best way to maintain beams with constant output characteristics, i.e., similar velocity and spatial distributions, is to divide the decelerator into a sequence of acceleration coils followed by a sequence of deceleration coils.  By using a constant value for the magnitude of the mean acceleration/deceleration, but varying the relative number of coils for acceleration and deceleration, one obtains beams with different mean velocities but constant phase-space acceptance, which is very useful for collision experiments.

(3) For a conventional Zeeman decelerator, it can be assumed that the timescale of the transverse dynamics is slow compared to the longitudinal dynamics, because both the transverse velocity and the transverse field gradients are lower compared to the longitudinal direction. We show that this adiabatic behavior is a good approximation in general, and allows for an explanation of the unfilled phase-space area for low amounts of acceleration/deceleration. However, we also demonstrate (by comparison with full 3D trajectory simulations) that such adiabatic calculations fail to account for certain empty regions of the phase-space-acceptance volume near to the boundaries marking the limits of stable trajectories. We conclude that non-adiabatic effects can occur under conditions where the period of the transverse dynamics and the longitudinal dynamics become similar and this coupling of longitudinal and transverse motion potentially leads to unstable trajectories. Although Wiederkehr et al. \cite{Wiederkehr2010} also pointed to this conclusion, the current paper provides substantial evidence for this, and therefore brings a greater understanding of phase stability in such devices. In practical terms, this leads to the conclusion that this non-adiabatic situation should be avoided if one wants to have beams with uniform spatial and velocity distributions.

(4) The paper shows that an adaptation of the '$s=1$' and '$s=3$' switching schemes, used elsewhere for Stark decelerators, is not particularly suitable for Zeeman deceleration. This is not {\it a priori} clear when considering only the longitudinal motion, as the shape of the longitudinal fields would look similar both in a Stark and a Zeeman decelerator; the differences therefore come as a consequence of the different transverse field curvature in Stark and Zeeman decelerators. Hence, this comparison of the different phase-space characteristics for Zeeman and Stark decelerators advances the general understanding of the principles of such devices and their design.

(5) More generally, our phase-space calculations for a Zeeman decelerator show that viable deceleration schemes can be developed even in the case of finite rise and fall times.  This fact is potentially very useful for the development of Stark decelerators, for which similar schemes should be applicable, since it has generally been assumed previously that very fast rise and fall times (and hence very expensive electronics) are required to guarantee phase stability. In addition, the concepts developed here are likely to be valuable for applications in the growing field of microwave deceleration in (superconductive) high finesse cavities, where switching on and off the fields rapidly is hampered due to the long lifetime of the photons.

\begin{acknowledgments}

We are grateful for numerous fruitful discussions with Michael Motsch (ETH Z\"{u}rich) and Bas van de Meerakker (Nijmegen University) that inspired some of the ideas presented in this article; and we thank Fr\'{e}d\'{e}ric Merkt (ETH Z\"{u}rich) for his ongoing interest in this work. K.D. acknowledges financial support from the Chemical Industry Fund (FCI, Germany) through a Kekul\'{e} Mobility Fellowship, and through a grant from the Simms Foundation (Merton College, Oxford, UK). This work is financed by the Engineering and Physical Sciences Research Council (U.K.) EPSRC(GB) under Project Nos. EP/G00224X/1 and EP/I029109/1.

\end{acknowledgments}



\end{document}